# THIRD INTEGER RESONANCE SLOW EXTRACTION SCHEME FOR A MU->E EXPERIMENT AT FERMILAB*

V. Nagaslaev#, J. Amundson, J. Johnstone, L. Michelotti, C.S. Park, S. Werkema,
FNAL, Batavia, IL 60510, U.S.A.
M. Syphers, MSU, East Lansing, MI 48825, U.S.A.

*Abstract*

The current design of beam preparation for a proposed mu->e conversion experiment at Fermilab is based on slow resonant extraction of protons from the Debuncher. The Debuncher ring will have to operate with beam intensities of $3 \times 10^{12}$ particles, approximately four orders of magnitude larger than its current value. The most challenging requirements on the beam quality are the spill uniformity and low losses in the presence of large space charge and momentum spread. We present results from simulations of third integer resonance extraction assisted by RF knock-out (RFKO), a technique developed for medical accelerators. Tune spreads up to 0.05 have been considered.

## INTRODUCTION

The Mu2e experiment, proposed at FNAL is aimed to search for rare neutrinoless decays of a muon to electron in the Coulomb field of the atomic nucleus [1]. This experiment is designed to be sensitive to muon conversion at the level of $6 \times 10^{-17}$, which improves existing experimental limits by 4 orders of magnitude. This requires a large suppression of the background. A pulsed structure of the proton beam suits this purpose. A veto gate allows prompt beam background to die down during 750ns, after which the detector is activated to look for mu-atom decays. The search time is limited by the muon lifetime in the atom (864ns), therefore, the time structure defined by the revolution time in the Debuncher, 1.69μs, is almost ideal for this scheme. A single bunch of 20-40ns width is formed in the Debuncher and resonantly extracted towards the mu2e production target. This bunch structure provides a substantial natural initial background suppression. Additional suppression is provided by external extinction system at level of $10^{-10}$. There are currently two alternative schemes of the resonant extraction under consideration: the half-integer and the third-integer resonance. Here we consider the latter one.

## DEBUNCHER

The 8 GeV proton beam from the FNAL Booster is sent to the Accumulator via the Recycler. Three batches of 53MHz Booster beam are momentum stacked and then rebunched into an h=4, 2.5MHz rf. Beam bunches then are sequentially transferred one at a time to the Debuncher and slowly extracted during 160ms.

The Debuncher ring has 3-fold symmetry, 3 arcs and 3 straight sections. In addition, each arc and each straight are mirror symmetric, giving the machine an overall dihedral symmetry. Presently the machine optics is not quite symmetric, in order to accommodate stochastic cooling and maximize the machine acceptance. After completion of Run-II, stochastic cooling equipment will be removed and the lattice symmetry will be restored to allow high intensity proton operation.

## SPACE CHARGE

Main requirements to the resonant extraction are the spill uniformity and minimal beam losses in the presence of substantial space charge tune shift. Slow spill from the Debuncher is done with a single bunch with rms length of 40ns and initial intensity of $3 \times 10^{12}$ protons. Space charge tune shift is therefore significant. Due to high dispersion in the arcs (Dx=2m) and finite momentum spread ($\sigma_p/p$= 0.004), however, this tune shift is reduced to about 0.015. It is very important for the experiment to keep the bunch length as low as possible. However, reduction of rms length down from 40ns requires a considerable increase of the rf power in the Debuncher, and therefore its cost. If the trade-off between cost and performance is made in favour of the latter, the bunch length will be reduced to 20-30ns, therefore increasing the space charge tune shift to 0.025-0.03. This kind of a tune spread with a strong asymmetry of the tune distribution represents difficulties for the resonant extraction, in particular when a good uniformity of the spill shape is required.

## TRACKING SIMULATIONS

Computer simulations of third-integer resonance extraction has been performed using the ORBIT code developed at ORNL [2]. Horizontal resonance tune was chosen at 29/3, the closest point to the current machine tune. Transfer matrices based on the improved symmetric lattice were used in this simulation. The sextupole field was formed by 2 orthogonal groups of 3 sextupoles, located in two straight sections. A quad circuit for tune ramping comprised 3 trim quads in the middle of each straight section. An extraction septum and lambertson magnet are located in the third straight. The septum width is assumed to be 100μ, as that used in other applications around the lab.

For calculating the space charge (SC) effects we used a so-called 2.5D-mode of ORBIT, where the particle density in longitudinal bins is calculated according to the actual longitudinal distribution, and the transverse distribution is assumed to be the same along the bunch.

___
*Work supported by DOE under contract No. DE-AC02-07CH11359
#vnagasl@fnal.gov

ORBIT SC calculation allows parallelization and computations were done on the multi-node farm Heimdall. Figure 1 shows an example of phase space calculation without SC (a) and with the SC (b). It can be seen that in the presence of SC extraction occurs from different separatrices due to the tune spread.

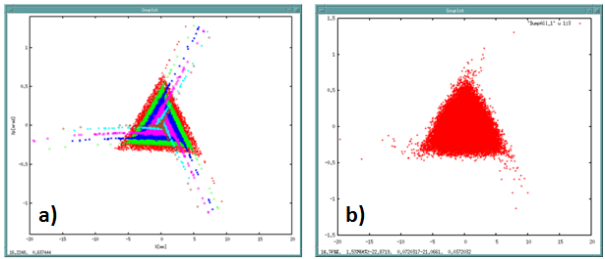

Figure 1. Normalized phase space of the beam: a) at different moments of extraction without SC; b) at the beginning of extraction with SC.

*Septum Losses*

Losses at the extraction point are a very essential concern of the project. The total beam power is 24kW, so losses at the level of 2% would produce 500W of radiation power localized release in the beam enclosure. Septum losses are determined by the ratio of its width and the step size (particle position increment after 3 turns). One of the advantages of the third integer resonance is that the step size is growing with the betatron amplitude and can be made large far from the separatix. Septum position and the sextupole field have been chosen to maximize the step size within limits of the machine acceptance.

*Ramps*

Two ramps need to be engaged in general to control the spill – the sextupole field ramp and the tune ramp. However, the sextupole field needs to be kept constant at the optimum setting, and also the spill has to start soon after injection, therefore the sextupole field ramps rapidly from zero to the nominal value and then stays constant. It will eventually be replaced with a constant field.

If the machine tune is set exactly to the resonance value, it takes a long time to extract all the beam, because particle tunes are separated from the resonance by the space charge tune shift, and those in the core of the beam are separated the most. We tried first approaching the resonance from above in order to expose those particles in the core the first. However, this approach increased losses on the septum, because in this case particles are extracted away from exact resonance and the step size is therefore reduced. Typically in this case septum losses start with about 5% in the beginning of the ramp and reduce to 1-2% in the end of the ramp. Approaching the resonance from below is more promising in terms of losses, although in this case it is more difficult to control extraction of the core particles in an uniform way. Figure 2a shows the distributions of the particle tune versus the horizontal action at the onset of the resonance.

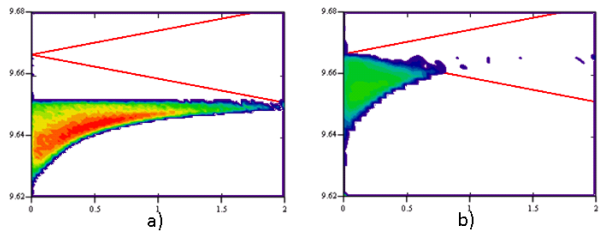

Figure 2. Tune distributions vs. horizontal action at the onset of the resonance and at exact resonance.

The starting machine tune is 9.650. Red lines show the 2/3 resonance extraction area boundaries due to the sextupole field. Figure 2b shows this distribution after machine tune been ramped to the exact resonance. Substantial part of the beam is still there and far from the resonance. After the tune ramp stopped at this point, extraction continues and the tune spread shrinks, which helps the extraction rate, but this rate is still very low. Extraction can be assisted with continuing the tune ramp and exercising multiple resonance crossing, but it is hard to control the spill rate uniformity in this case.

## RF KNOCK-OUT

We propose another way to assist extraction in the situation of Figure 2b. If one succeeds to heat the beam transversely fast enough, the tune distribution would move to the right and up. This would make it closer to the extraction area on one hand, and with a proper mixing, would also reduce the SC tune spread. Such a technique already has been used for slow extraction purposes in medical applications [3], although the primary goal of that was to turn off/on the beam extraction. This technique is known as RF knock-out (RFKO). We are using RFKO as a feed-back tool for the fine control of the spill rate.

RFKO allows us to continue extraction with presence of the strong space charge while keeping the machine tune close to the resonance. In this case particles are extracted on the resonance, therefore the step size is maximized.

Sufficient transverse E-field may be provided by a regular damper. Frequency modulation around a single betatron sideband is required to sweep the excitation frequency within the tune spread of the beam. Coloured noise modulation (random signal within a given bandwidth) appears to be the best way of modulation, however we didn't find its advantages in performance compared to the normal sweeping with phase randomization between sweeps, so we used the latter approach in our simulations. Figure 3 shows the RFKO power (red) and instant spill (green) rate during the spill. The dashed blue line shows the nominal rate.

RFKO power is used here as a feedback knob to control the spill uniformity. Although sophisticated and intelligent techniques exist to manage feedbacks, a simple filter was used in simulations. The power setting was updated once every 100 turns. If the extraction rate is higher than nominal at the moment of update, power is reduced by a constant down-factor. Power is multiplied by an up-factor only if the rate is lower than nominal and not growing. Normally it takes about 1ms for the feedback to take effect, therefore the up-factor is chosen to be above and close to 1.0. When the rate started to grow, it grows fast, therefore the down-factor should be small. One can see in Figure 3 that RFKO power is growing slowly and drops down fast due to this choice of u- and d-factors. Maximum RFKO power has been limited by available hardware specs. Instantaneous rate variations are substantial but hard to avoid. However, what matters most for the experiment is the integrated uniformity. As shown in Figure 4d, the overall beam intensity curve looks good. Figures 4 a)-c) show respectively the quad, tune and sextupole ramps during the spill.

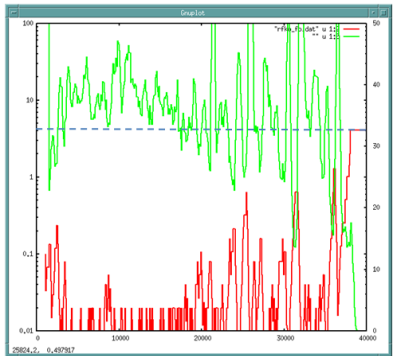

Figure 3. Instant spill rate (green) and RFKO power (log scale) applied to correct the spill rate.

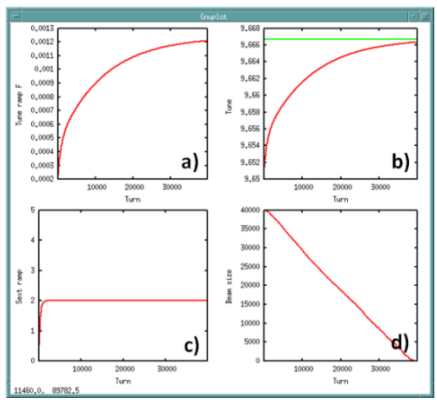

Figure 4. Quad circuit ramp (a), tune (b), sextupole (c) and beam intensity (d) during the spill. Green line in (b) shows 2/3 resonance. All units are arbitrary except of (b).

It is worth noting that the effectiveness of the beam heating with RFKO is limited, so the tune ramp curve (as Figure 4b) should be chosen carefully to facilitate extraction as close to constant rate as possible without RFKO.

Losses on the septum defined as a ratio of a number of particles that have hit the septum wire to the total number of extracted particles is typically around 2% or better with this scheme of extraction, and there is no obvious dependence on the time in spill. Numeric computations of geometric inefficiency were also made based on the analytical calculations in conditions close to those we used in our simulations, although not including the space charge. The dark curve in Figure 5 represents those computations. Based on those calculations the simulations have been performed [4] using tools other than ORBIT, their results are shown as red dots in Figure 5.

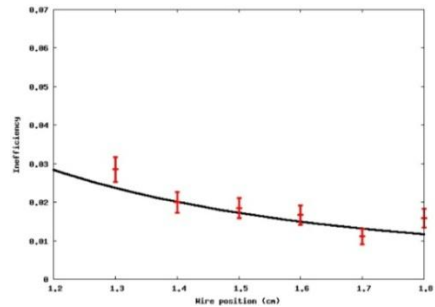

Figure 5. Numeric computations of the geometric septum losses versus the septum position.

Nominal septum position is at 1.4cm, which corresponds to inefficiency about 2% that is very close to our observations.

## SUMMARY

We proposed a method of controlling the spill rate in the presence of a strong space charge tune shift. It has been shown in the tracking simulations using the ORBIT package that this can effectively help to make spill uniform. Simulations were also extended to higher intensities and tune shifts up to 0.05 and similar results have been obtained. With realistic beam and lattice parameters it was shown that septum losses can be kept at or below 2%. Analytic calculations were made for the extraction inefficiency and results obtained are in a good agreement with the simulations results.

## AKNOWLEDGEMENTS

We are grateful to Dave Wildman for his help in identifying RFKO hardware components and Oleksii Nikulkov for his assistance in simulating data on inefficiency without SC effects.

## REFERENCES

[1] Mu2e Proposal, Document 388-V1 at http://mu2e-docdb.fnal.gov.
[2] J.Galambos et al., "ORBIT-A Ring Injection Code with Space Charge", PAC-99.
[3] M.Tomizada et al., "Slow Beam Extraction at TARN II", NIM A326 (93) 399; K.Noda et al., "Slow Beam Extraction by a Transverse RF Field with AM and FM", NIM A374 (96) 269.
[4] L.Michelotti, J.Johnstone, "Preliminaries toward studying resonant extraction from the Debuncher", FERMILAB-FN-0842-APC-CD; L.Michelotti, "Step size, efficiency, and the septum; notes from quadrature", mu2e document 1021, http://mu2e-docdb.fnal.gov.